\documentclass[twocolumn,fleqn,a4paper]{article} 
\usepackage{epsfig}
\usepackage[english]{babel}
\begin{document}

\title{Jet-Studies and $\alpha_s$-determinations at HERA\footnote{Talk given on
behalf of the H1- and ZEUS-Collaborations at the conference "New Trends in
High-Energy Physics", Alushta, Crimea, Ukraine, May 24-31, 2003}.}

\author{Gerd W. Buschhorn \\ Max-Planck-Institut f\"ur
Physik \\(Werner-Heisenberg-Institut) \\F\"ohringer Ring 6 \\80805 M\"unchen} 
\date{}
\maketitle

\begin{abstract}Recent results from the H1 and ZEUS Collaborations on inclusive
single and multiple jet production in the neutral current deep inelastic
scattering of electrons/positrons on protons and the high energy
photoproduction on protons at HERA are reported. The results are compared with
NLO QCD calculations and have been used to determine the strong interaction
coupling constant $\alpha_s$.
\end{abstract}

\section{Introduction}

The study of jets in the deep inelastic scattering (DIS) of electrons/positrons
on protons and in high energy photoproduction on protons enables sensitive
tests of QCD to NLO and provides precision measurements of the strong
interaction coupling constant $\alpha_s$. In this report a survey is given of
results from recent measurements of jet production in neutral current processes
from the H1 and the ZEUS Collaborations at the HERA collider; not included in
this review are results on jets associated with diffraction, heavy flavours or
direct photons.

\begin{figure}[ht]
\begin{center}
\epsfig{file=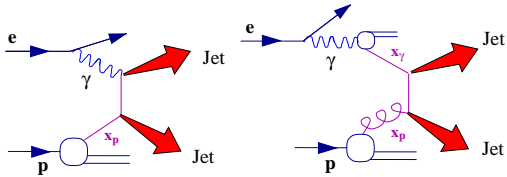,width=7.5cm}
\caption{\small Examples of leading-order contributions to
inclusive jet production: (left) direct photon interaction, (right) resolved
photon interaction.}
\label{fig1}
\end{center}
\end{figure}

In jet studies QCD predictions for DIS and photoproduction are probed in
different regions of the DIS variables (Bjorken)$x$ and $Q^2$ (with $y =
Q^2/sx$ and $s$ the squared $ep$ centre-of-mass energy) and the transverse
energy $E_{\rm T}$ and pseudorapidity $\eta_j$ (defined as $\eta_j = - \log
\rm{tg} \hspace{1mm} \Theta_j/2$ with $\Theta_j$ the polar jet angle with
respect to the proton beam) of the hadronic jet or jets. In DIS at high $Q^2$
the virtual photon emitted from the electron interacts with a parton from the
proton in a direct (pointlike) process while at low $Q^2$ and in
photoproduction it can fluctuate into partons, one of which interacts in
so-called resolved processes with a parton of the proton (fig.~\ref{fig1}).

\begin{figure}[ht]
\begin{center}
\epsfig{file=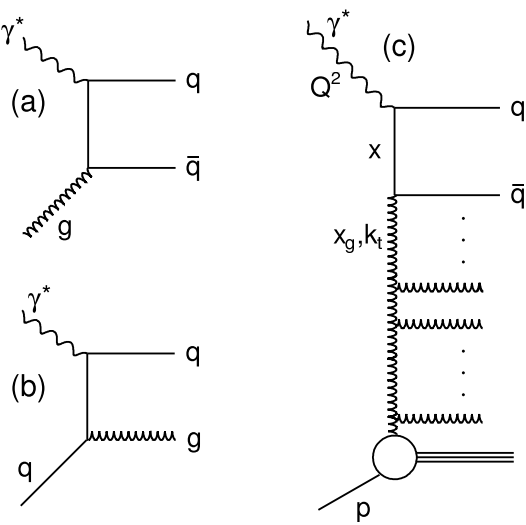,width=7cm}
\caption{\small Leading order diagrams for (di)jet production:
(a) photon-gluon fusion; (b) QCD-Compton process; (c) parton cascade.}
\label{fig2}
\end{center}
\end{figure}

In pQCD jet production depends on the hard partonic scattering process and on
the parton density functions (PDFs) of the proton and, in photoproduction and
low $Q^2$ processes, also of the photon. For the calculation of the jet cross
section $\sigma_j$ in DIS (for photoproduction see Chap.~3), the hard partonic
cross section  $d\sigma_a$ for a parton $a$ has to be convoluted with the
PDFs $f_a$ of the proton

\begin{eqnarray}  
\sigma_j & = & \sum_a \int dxf_a(x,\mu^2_F) d\sigma_a(x,\alpha_s(\mu^2_R),
\mu^2_R, \mu^2_F) \nonumber \\ 
& & \cdot (1 + \delta_{\rm hadr}) 
\end{eqnarray}

\noindent and corrected for hadronization effects. The dependence of the PDFs
on the characteristic energy scale of the interaction is described by evolution
equations i.e. for not too small $x$ and $Q^2$ by the DGLAP equations. The
summed higher order contributions of this evolution, corresponding to multiple
collinear gluon emissions of the interacting parton, are represented by parton
cascade diagrams. The factorization theorem allows the soft (nonperturbative)
part of these contributions to be absorbed in the PDFs while the hard part has
to be accounted for in the perturbative part of the interaction, it defines the
factorization scale $\mu_F$. Examples of LO contributions to DIS are shown in
fig.~\ref{fig2}. A breakdown of the DGLAP approximation is to be expected in
the small $x$ region, where $\log 1/x$-terms become larger than $\log
Q^2$-terms; in this region, the BFKL evolution equation should give a better
description of DIS. A kind of interpolation between these approximations is
provided by the CCFM approach.

A preferred reference system for analyzing jet processes in DIS is the Breit
system in which the parton collides head-on with the timelike virtual photon:
$2x$\boldmath$p$\unboldmath$+$\boldmath$q$\unboldmath$=0$. In the quark parton
model (QPM) the hadron jet resulting from the backscattered quark emerges
collinear with the proton remnant, whereas QCD hard processes are characterized
by jets with high transverse momenta with respect to the proton remnants.
Analyzing the hadronic final state in the Breit system therefore suppresses the
QPM background.

The identification of jets is achieved by algorithms which are applied to the
objects of the measurement i.e. the particle tracks and energy clusters that
are registered by the detector. Different jet algorithms have been developed
and applied to the HERA experiments but in recent years the "inclusive
longitudinally invariant $k_{\rm T}$ cluster algorithm" (in the following:
ILICA) has become the preferred one. The algorithm proceeds by calculating for
each track or energy cluster $i$ the quantity $d_i$ and for each pair the
quantity $d_{ij}$:

\begin{eqnarray}  
d_i & = & E_{{\rm T},i}^2, \nonumber\\
d_{ij} & = & \min(E_{{\rm T},i}^2, E_{{\rm T},j}^2)(\Delta \eta_{i,j}^2 +
\Delta \phi_{i,j}^2)
\end{eqnarray}

\noindent where $E_{{\rm T},i}$ is the transverse energy of a particle $i$ and
$\Delta \eta_{ij}$ and $\Delta \phi_{ij}$ are the differences in the
pseudorapidity $\eta$ and azimuthal angle $\phi$ of $i,j$ respectively; all
these quantities are invariant under longitudinal boosts i.e. boosts in the
direction of the proton beam (defined as $+z$-direction). If of all resulting
\{$d_i, d_{ij}$\} the smallest one belongs to \{$d_i$\} it is kept as a jet and
not treated further if it belongs to \{$d_{ij}$\} the corresponding tracks or
clusters are combined to a single object. The procedure is repeated until all
tracks or clusters are assigned to jets. The algorithm is infrared and
collinear safe, and can be applied to measurements of DIS and photoproduction
processes. It typically requires smaller hadronization corrections than other
clustering algorithms.

For tests of QCD, the measured jet cross sections have to be compared with
calculations performed (at least) to NLO.  These partonic final states have to
be combined with models for the development of the parton cascade and the
hadronization. Models for the parton cascade are the parton shower model,
implemented in HERWIG and LEPTO, and the dipole cascade model, implemented in
ARIADNE; the hadronization of the partons is modelled by the cluster model,
implemented in HERWIG, or by the string fragmentation model, implemented in
JETSET and used in LEPTO and ARIADNE. In RAPGAP the LO QCD matrix elements are
matched to DGLAP based parton showers in the LL approximation; it allows
RAPGAP, besides simulating direct processes, also the simulation of resolved
photon processes. In CASCADE, the QCD matrix elements are combined with parton
emissions  described by the CCFM evolution equation with the unintegrated gluon
density distribution used as input.

In photoproduction, direct and resolved interactions are simulated with PYTHIA,
HERWIG and PHOJET. The hard partonic interaction is described by LO QCD matrix
elements and the parton cascade is simulated by initial and final state parton
showers in the LL approximation. In PYTHIA and PHOJET, the hadronization is
simulated with the Lund string model as implemented in JETSET; in HERWIG the
cluster model is used.

Commonly used parametrizations for the PDFs of the proton are those of CTEQ and
MRST, while for the photon PDFs the GRV and AFG sets are used. 

\section{Jets in deep inelastic scattering}

Results on inclusive jet production in DIS have been obtained on single jets,
dijets and trijets and recently also on subjets. Since jet production at small
$x$ in the forward direction is related to forward single hadron production,
results on $\pi^{\circ}$-production are also discussed.

\subsection{Inclusive jets}

\begin{figure}[ht]
\begin{center}
\epsfig{file=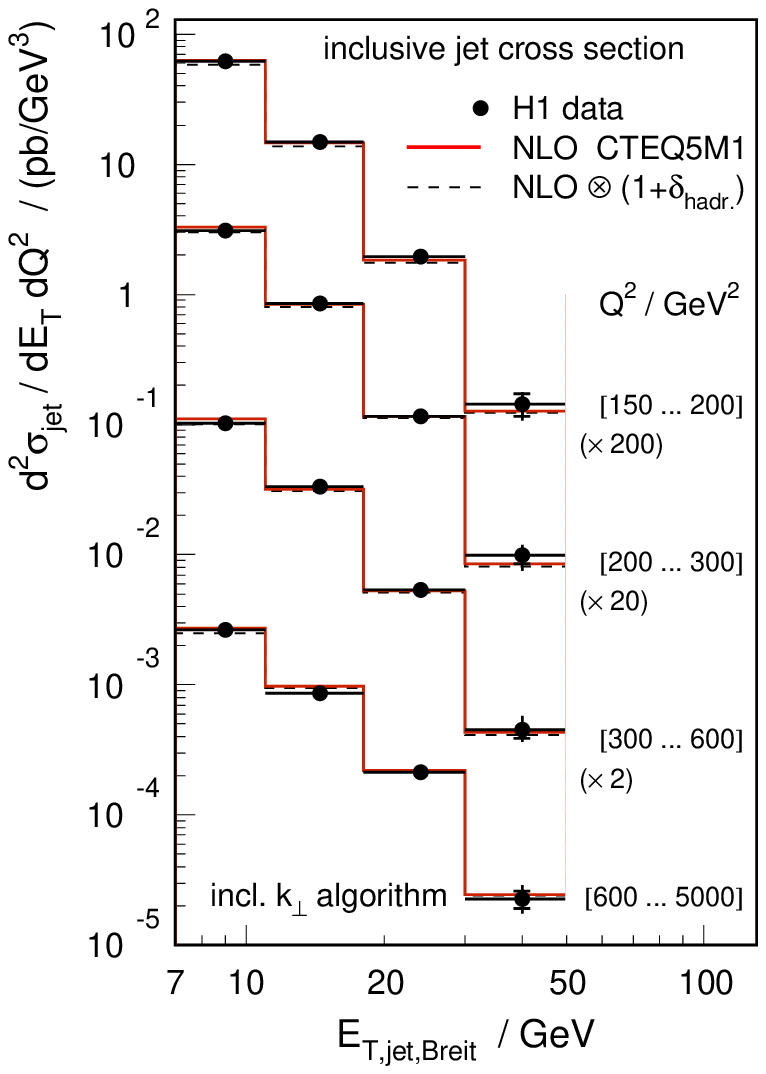,width=7cm}
\caption{\small Inclusive jet cross section in DIS as fct. of the transverse
jet energy in the Breit system $E_{{\rm T}jB}$.}
\label{fig3}
\end{center}
\end{figure}

At high $Q^2$, i.e. for $150$ GeV$^2 < Q^2 < 5000$ GeV$^2$, the inclusive jet
cross section has been measured by H1 \cite{r1} in the Breit frame as a
function of the jet transverse energy $E_{\rm T}$ for 7 GeV $< E_{\rm T} < 50$
GeV, $0.2 < y < 0.6$ and $-1 < \eta_{\rm{ lab}} < 2.5.$ Jets were
identified using the ILICA. Over the whole range of $E_{\rm T}$ and $Q^2$, the
NLO calculation corrected for hadronization effects ($< 10$\%) gives a good
description of the data (fig.~\ref{fig3}).

Fitting the QCD prediction to these cross sections using CTEQ5M1, $\mu_R =
E_{\rm T}$ and $\mu_F = \sqrt{200\hspace{2pt}{\rm GeV}^2}$ (the average $E_{\rm
T}$ of the jet sample) yields $\alpha_s$. First in separate fits to the data
points in the four $Q^2$ regions $\alpha_s (E_{\rm T})$ was determined and
evolved to $\alpha_s (M_Z)$. The consistency of these fits justified a combined
fit to all data points with the final result, taking into account the
uncertainties of the renormalization scale $(\mu_R = E_{Tj})$ and from the
parton parametrization, of $\alpha_s (M_Z) = 0.1186 \pm 0.0030 \rm{(exp)}
+0.0039/-0.0045 \rm{(theor)} +0.0033/-0.0023$ (PDF). The experimental error is
dominated by the hadronic energy scale of the LAr calorimeter with the
statistical error contributing only 0.0007; the theoretical uncertainty
includes about equal contributions from the hadronization and renormalization
scale uncertainty. The result has been shown to be stable against variations of
the jet algorithm.

Similar measurements have been performed recently by ZEUS \cite{r2}. For $Q^2 >
125$ GeV$^2, E_{{\rm T}B} > 8$ GeV, $-2 < \eta_{jB} < 1.8$ and demanding for
the angle $\gamma$ of the hadronic system $-0.7 < \cos_{\gamma} < 0.5$, the
$Q^2$- and $E_{{\rm T}B}$-dependence of the inclusive jet cross section is
found to be in good agreement with NLO QCD calculations.

For the determination of $\alpha_s$, the differential cross sections
$d\sigma/dV$, where $V = Q^2, E_{{\rm T}Bj}$, were calculated in NLO QCD for
three MRST99 sets i.e. central, $\alpha_s \uparrow \uparrow$ and $\alpha_s
\downarrow \downarrow$, taking for the partonic cross sections the $\alpha_s
(M_Z)$ of the corresponding PDFs. For each bin $i$ in $V$, the calculated cross
sections were parametrized according to $[d\sigma(\alpha_s(M_Z))/dV]_i =
C_{1,i} \cdot \alpha_s (M_Z) + C_{2,i} \cdot\alpha_s^2 (M_Z)$. The best
combined fit ($\chi^2 = 2.1$ for 4 data points) was achieved for $Q^2 > 500$
GeV$^2$ resulting in $\alpha_s (M_Z) = 0.1212 \pm 0.0017$ (stat) $+0.0034 \atop
-0.0045$ of the data.

\begin{figure}[ht] 
\begin{center} 
\epsfig{file=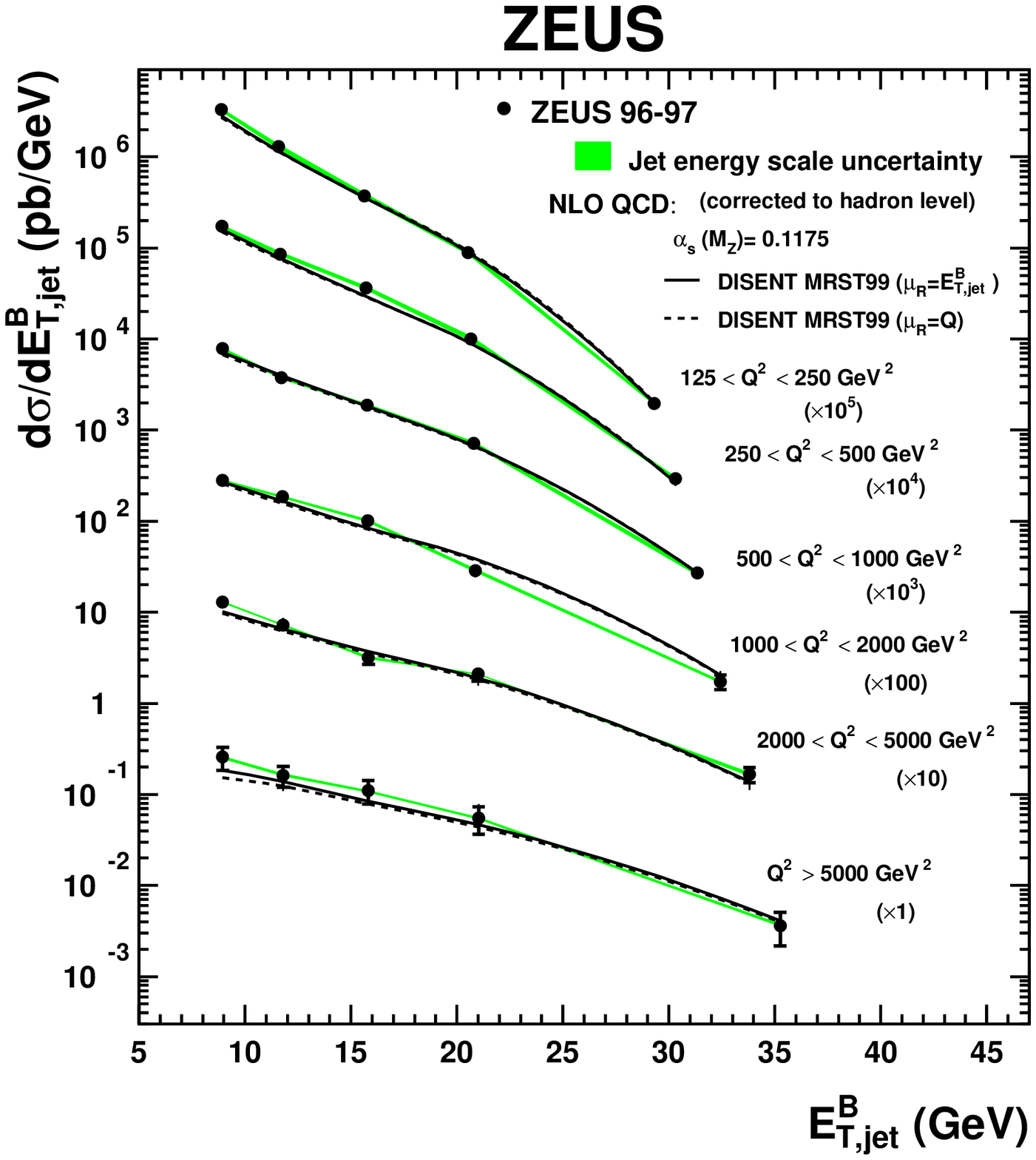,width=7.5cm}
\caption{\small Inclusive jet cross section in DIS as fct. of the transverse
jet energy in the Breit system $E_{{\rm T}jB}$.}
\label{fig4} 
\end{center} 
\end{figure}

The energy scale dependence of $\alpha_s$ has been investigated using the same
procedure but parametrizing the measured cross section in terms of
$\alpha_s(\langle E_{\rm T_{jet}}^B \rangle)$, where $\langle E_{\rm T_{jet}}^B
\rangle$ ist the mean value of $E_{\rm T_{jet}}^B$ in each bin; $\mu_R = E_{\rm
T_{jet}}^B$ was chosen here since it provides the better description of these
data. Good agreement with the predicted running of $\alpha_s$ is found over a
large range of $E_{\rm T_{jet}}^B$.

\begin{figure}[ht] 
\begin{center} 
\epsfig{file=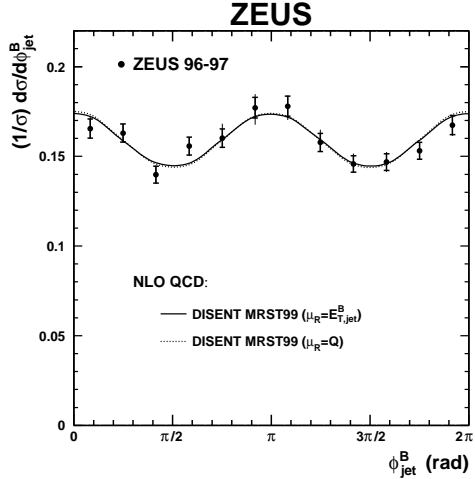,width=7.5cm}
\caption{\small Normalized differential cross section for inclusive jet
production in DIS with $E_{{\rm T}B_j} > 8$ GeV and $-2 < \eta_{B_j} < 1.8$;
inner errors for statistical, outer errors for combined statistical and
systematic errors.} 
\label{fig5}
\end{center} 
\end{figure}

These data have also been used to study the azimuthal distribution of jets in
the Breit frame \cite{r3}. At LO, the dependence of the cross section on the
angle $\phi_j^B$ between the lepton scattering plane and jets produced with
high $E_{\rm T}$ takes the form

\begin{equation} 
\frac{d\sigma}{d\phi^B_j} = A + B \cos \phi^B_j + C \cos 2\phi_j^B
\end{equation}

\noindent where the coefficients $A, B, C$ result from the convolution of the
matrix elements for the partonic processes with the PDFs of the proton. The
$\cos \phi$-term arises from the interference between the transverse and
longitudinal components of the exchanged photon (for $Q^2 \ll M_Z^2$) whereas
the $\cos 2\phi$-term results from the interference of the +1/-1 -helicity
amplitudes of its transversely polarized part; nonperturbative contributions
arising from intrinsic transverse momenta of the partons have shown to be small
and are negligible at high $E_{\rm T}$. The measured normalized
azimuthal distribution shown in fig.~\ref{fig5} is well reproduced by NLO
calculations with either $\mu_R = E_{\rm T}$ or $Q$. 

\begin{figure}[ht]
\begin{center}
\epsfig{file=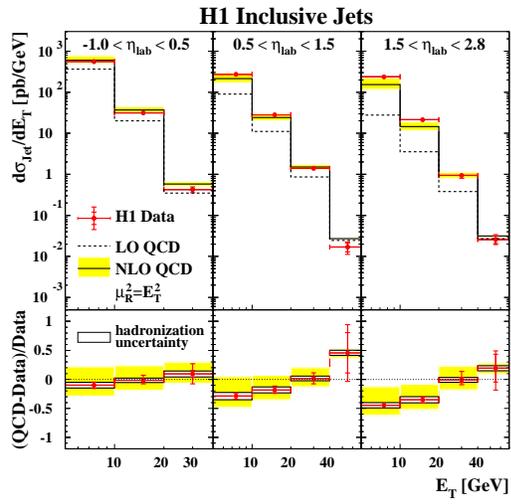,width=7cm}
\caption{\small Inclusive jet cross section $d\sigma/dE_{\rm T}$ in DIS
integrated over 5 GeV$^2 < Q^2 < 100$ GeV$^2$ and $0.2 < \eta_{\rm lab} < 0.6$
as function of the transverse jet energy $E_{\rm T}$ in the Breit system for
different ranges of $\eta_{\rm lab}$: forward (proton direction): $1.5 < \eta <
2.8$; central: $0.5 < \eta < 1.5$; backward: $-1.0 < \eta < 0.5$. NLO QCD
model: DISENT with CTEQ5M (solid line) and no hadronization corrections;
hatched band: change of the renormalization scale by 1/2 and 2.}
\label{fig6}
\end{center}
\end{figure}

H1 has extended the inclusive jet measurements to low $Q^2$ values of 5 GeV$^2
< Q^2 < 100$ GeV$^2$ for $0.2 < y < 0.6$, $E_{\rm T} > 5$ GeV and $-1 <
\eta_{\rm lab} < 2.8$ \cite{r4}. The differential cross section as a function
of $E_{\rm T}$ is shown in fig.~\ref{fig6} for three bins of $\eta$. The good
agreement with NLO QCD predictions (for $\mu_R = E_{\rm T}$) found for the
backward and central region is getting worse in the forward region i.e. of 
$\eta > 1.5$, and for $E_{\rm T} < 20$ GeV, where NLO corrections and
uncertainties are becoming larger. For most of the data the theoretical
uncertainties (given by the renormalization scale uncertainty) are larger that
the experimental errors. Investigating the $Q^2$ dependence of this
discrepancy, it was found to originate from the $Q^2 < 20$ GeV$^2$ region,
indicating that NLO QCD works reasonably well even in the forward region
$\eta_{\rm lab} > 1.5$ as long as both $E_{\rm T}$ and $Q^2$ are not too
small. 

\begin{figure}[p]
\begin{center}
\epsfig{file=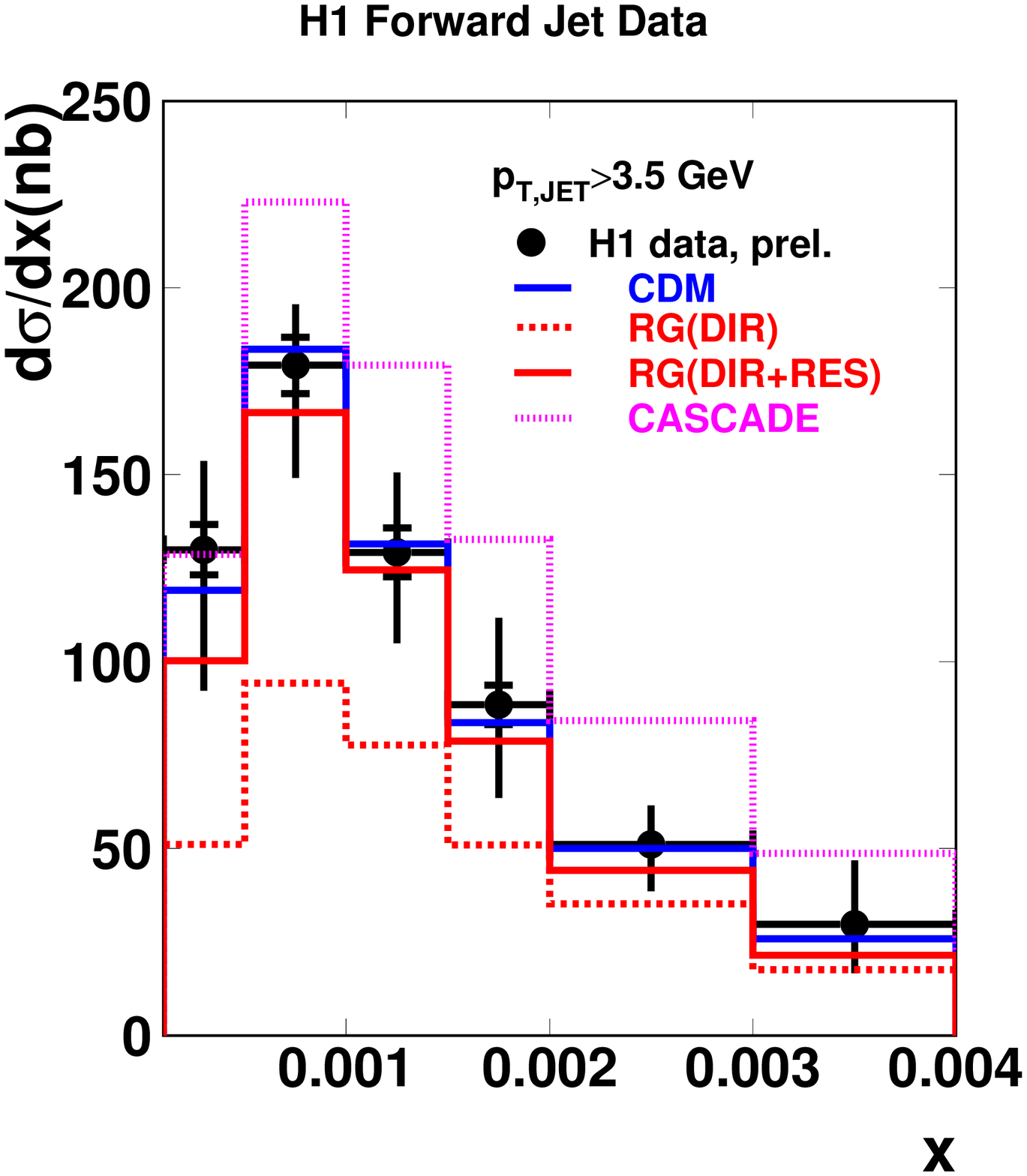,width=7.5cm}
\epsfig{file=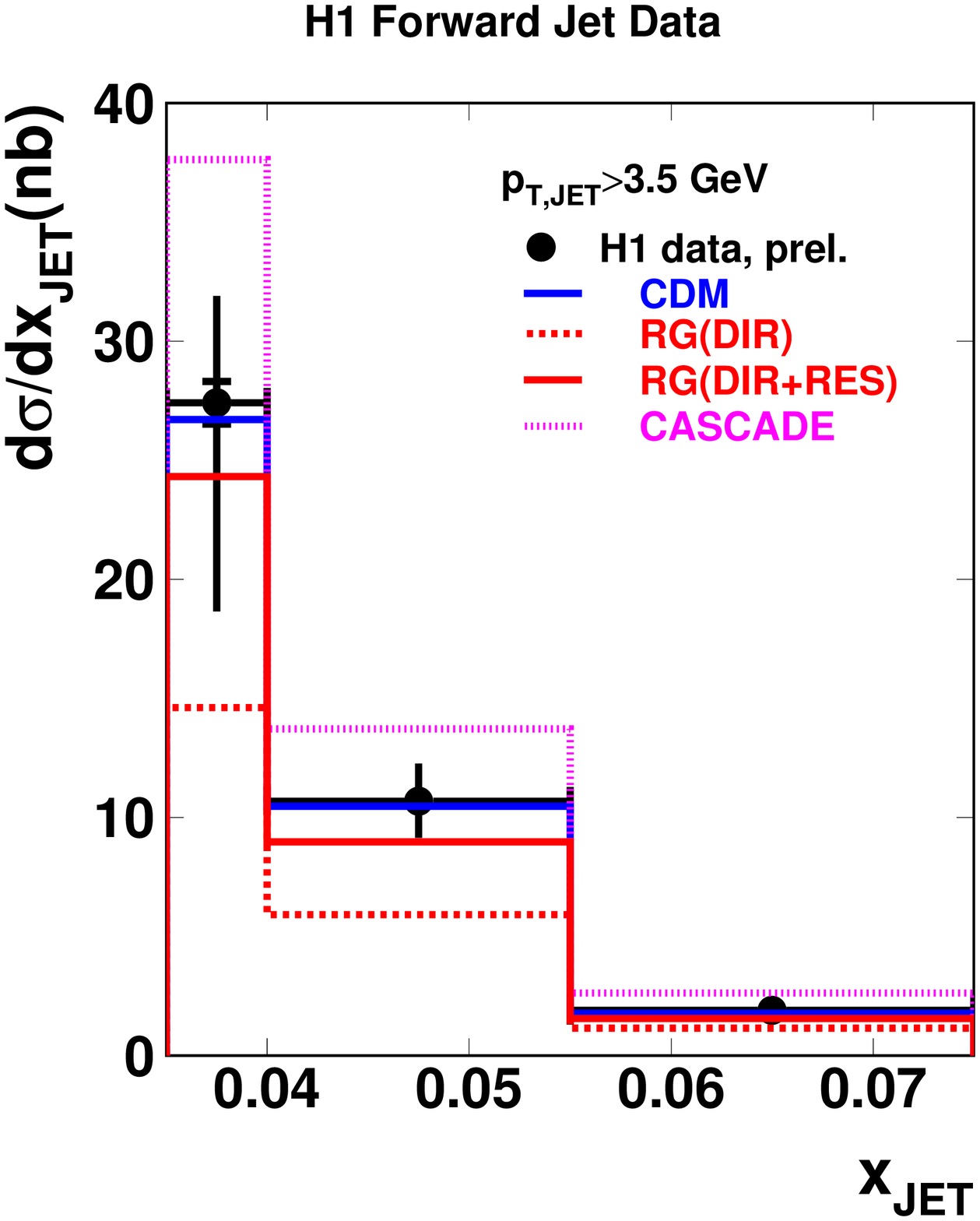,width=7.5cm}
\caption{\small Inclusive forward jet cross sections in DIS for
$p_{tj} > 3.5$ GeV; $7^{\circ} < \Theta_j < 20^{\circ}; x_j > 0.035$ at the
hadron level. Models: CDM (Color-Dipole-Model: DJANGO + ARIADNE), RG (RAPGAP:
DIRECT and DIRECT + RESOLVED), CASCADE (CCFM). Data are preliminary.}
\label{fig7}
\end{center}
\end{figure}

In the forward direction, jet production is expected to be sensitive to the
dynamics of the parton cascade at small $x$. As suggested by Mueller-Navelet,
the contribution of DGLAP evolution to DIS should be suppressed by demanding of
the forward jet $E_{\rm T}^2 \approx Q^2$ while that of BFKL evolution should
be enhanced by keeping $x_j$ as large as feasible and $x/x_j$ small.

H1 has taken data on inclusive jet production in the forward direction
\cite{r5} i.e. $7^{\circ} < \Theta_{j {\rm lab}} < 20^{\circ}$ with $0.5 <
Q^2/E_{\rm T}^2 < 2$ and $x_j >0.035$ ($E_{\rm T} > 3.5$ GeV). The measurement
and jet search with the ILICA is performed in the lab frame with cuts $5 < Q^2
< 75$ GeV$^2$ and $0.1 < y < 0.7$. The data (fig.~\ref{fig7}) are compared with
four different Monte Carlos. The differential cross sections in $x$ and $x_j$
are up to a factor of two larger than the DGLAP prediction for direct photons
only (RG (DIR)), but they are reasonably well described if a resolved photon is
included. The description provided by the CDM model is also good while CCFM
predicts too large cross sections.

\begin{figure}[ht]
\begin{center}
\epsfig{file=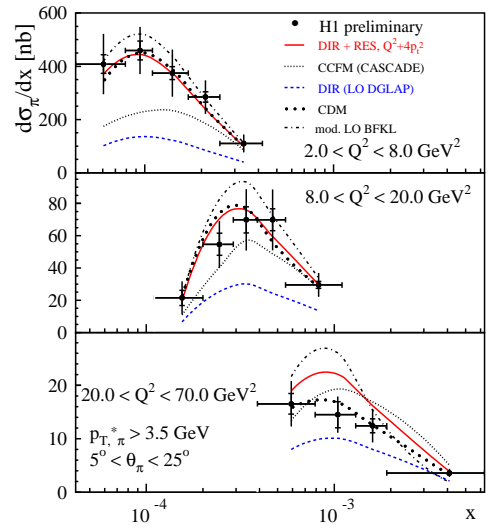,width=7.5cm}
\caption{\small Inclusive forward $\pi^{\circ}$ cross section
in DIS for transverse momenta in the $\gamma$-proton cms $p_{{\rm T} \pi^{\circ
}}^*  > 3.5$ GeV and $5^{\circ} < \Theta \pi^{\circ} < 25^{\circ}$ for
different ranges of $Q^2$. Models: DIR (RAPGAP for DIRECT proc.), DIR + RES
(RAPGAP for DIRECT + RESOLVED proc.), CCFM (CASCADE), CDM (Color-Dipole-Model:
ARIADNE), mod. LO BFKL (modified LO BFKL evolution).}
\label{fig8}
\end{center}
\end{figure}

In forward $\pi^{\circ}$ production, the parton is tagged by a single energetic
fragmentation product. While smaller forward angles than in jet production can
be reached, the cross sections are lower and the hadronization uncertainties
higher.

Inclusive $\pi^{\circ}$-production in DIS has been measured by H1 \cite{r6},
with the following cuts: $0.1 < y < 0.6$, $4.10^{-5} < x < 6.10^{-3}$, 2 GeV$^2
< Q^2 < 70$ GeV$^2$ and $\pi^{\circ}$ cms-momenta $> 2.5$ GeV, polar angles of
$5^{\circ} < \Theta_{\pi} < 25^{\circ}$ (corresponding to the central region in
the hadronic cms) and $x_{\pi} > 0.01$; no explicit cut on $p_{{\rm T}
{\pi^{\circ}}}^2/Q^2$ was applied. In fig.~\ref{fig8}, the differential cross
section is compared with QCD-based models. The DGLAP prediction with direct
photon interaction only is too low, but the inclusion of resolved photon
interaction - albeit with a choice of the renormalization and factorization
scale different ($Q^2 + 4p_{\rm T}^2$) than in forward jet production - yields
a good description. BFKL based calculation also describes the data reasonably
well, while CCFM evolution fails at small $x$.

\begin{figure}[p]
\begin{center}
\epsfig{file=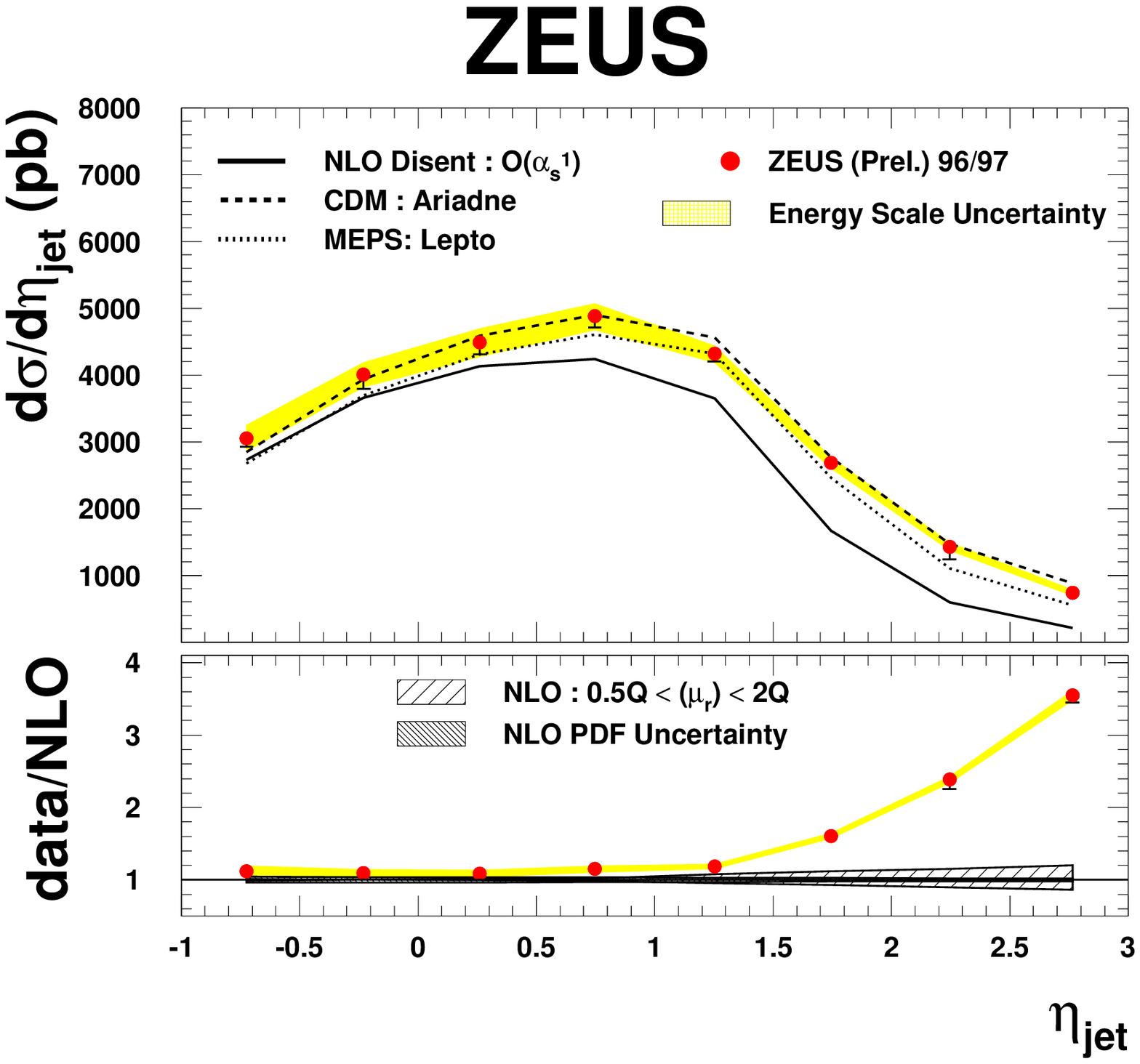,width=7.5cm}
\epsfig{file=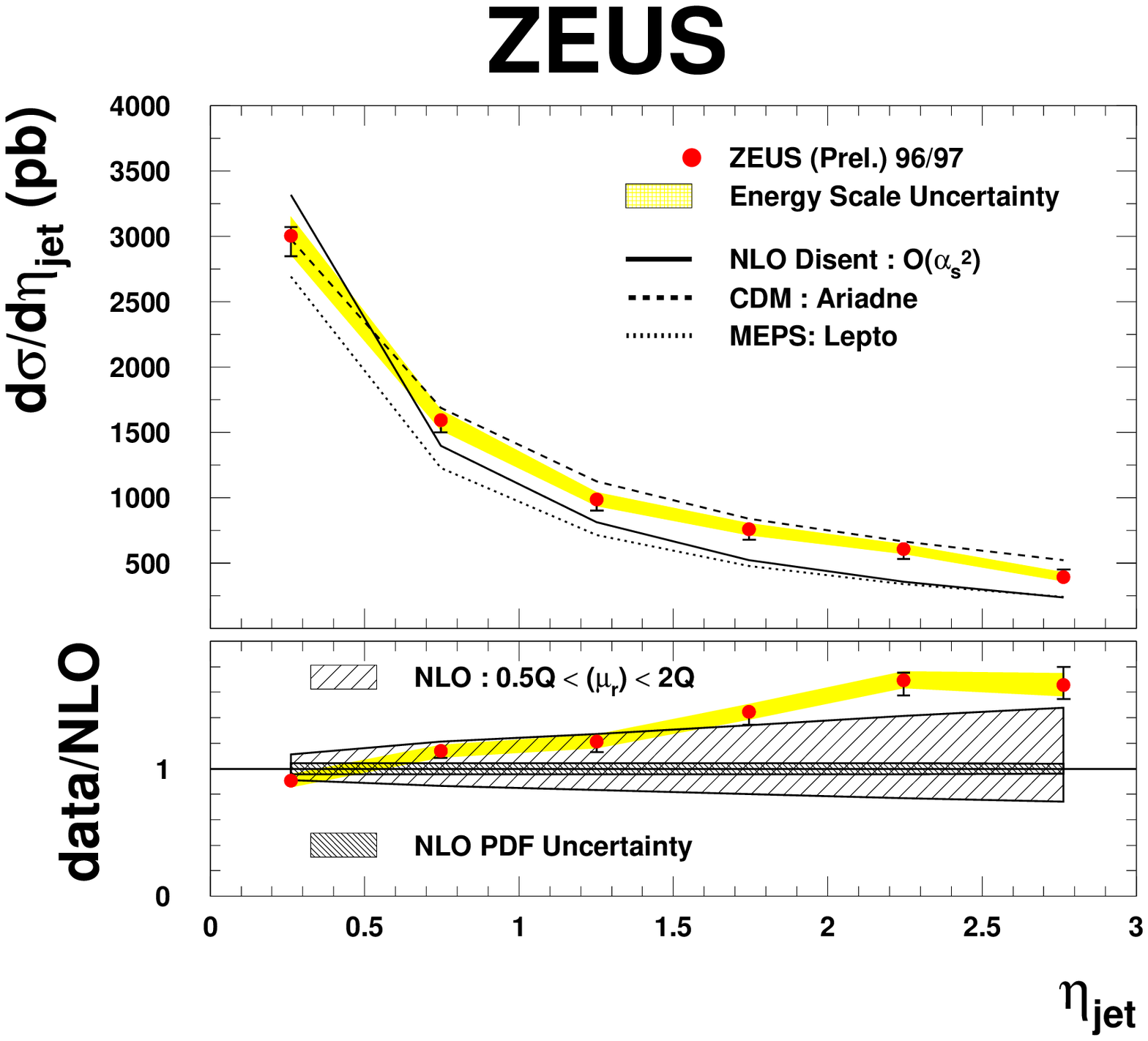,width=7.5cm}
\vskip-2cm
\caption{\small Upper figure: Inclusive jet cross sections in
DIS for $Q^2 > 25$ GeV$^2$, $y > 0.04$, $E_{{\rm T}j} > 6$ GeV and $-1 < \eta_j
< 3$ in comparison to $O(\alpha_s^1)$-NLO QCD calculations in the lab frame
("inclusive phase space", see text). Lower figure: Jet cross sections for the
same kinematical conditions but constraining the jet topology to $\eta_j > 0$
and $\cos \gamma_{hadr} < 0$ in comparison with $O(\alpha_s^2)$-NLO QCD
calculations in the lab frame ("dijet phase space").}
\label{fig9}
\end{center}
\end{figure}

In a recent measurement by ZEUS \cite{r7} of inclusive jet production in DIS
for $Q^2 > 25$ GeV$^2$, $y > 0.4$ and with $E_{{\rm T}j} > 6$ GeV, $-1 < \eta_j
< 3$, deviations of the data from NLO QCD predictions calculated in the lab
frame to $O(\alpha_s^1)$ where found to increase towards the forward direction
(fig.~\ref{fig9}, upper part) and small $E_{{\rm T}j}, Q^2$ and $x$. In a
further analysis of these data, the Mueller-Navelet suggestion was applied by
combining the detection of a hard forward jet ($E_{{\rm T}j}> 6$ GeV, $0 <
\eta_j < 3$) with requiring the hadronic angle $\gamma_{\rm hadr}$,
corresponding to the direction of the scattered quark in the QPM, to be
reconstructed in the backward direction of the detector ($\cos \gamma_{\rm
hadr} < 0$) and thus suppressing QPM contributions to the forward jets.
Comparing the cross sections for such defined jet configurations ("dijet phase
space") with NLO QCD predictions calculated to $O (\alpha_s^2)$, better
agreement (fig.~\ref{fig9}, lower part) is found in the forward direction at
the expense of a considerably larger uncertainty in the renormalization scale
which swamps a possible BFKL signal in this part of the phase space.

\subsection{Dijets}

The LO QCD processes which contribute to dijet production in DIS are QCD
Compton scattering (QCDC) and boson-gluon-fusion (BGF). In the
high $Q^2$ region, where the QCDC process is dominant and the PDFs are well
constrained by fits to inclusive DIS data, dijet measurements with $\alpha_s$
as input enable tests of pQCD; at low $Q^2$, where BGF dominates, the
comparison of measured dijet cross sections with QCD predictions for different
PDFs has been used to investigate the gluon density distribution.

\begin{figure}[ht]
\begin{center}
\epsfig{file=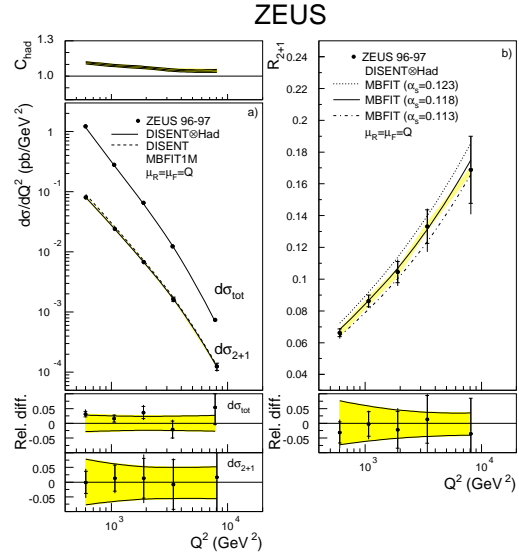,width=7.5cm}
\caption{\small (left)  Inclusive jet ($\sigma_{\rm tot}$) and
dijet ($\sigma_{2+1}$) cross sections in DIS, hadroniz. corr. $C$ refers to dijets;
(right): dijet fraction $R_{2+1}$, shaded band gives uncertainty of abs. energy
scale of jets.}
\label{fig10}
\end{center}
\end{figure}

In dijet measurements by H1 \cite{r1} and ZEUS \cite{r8} asymmetric cuts in
$E_{\rm T}$ have been applied to the two jets highest $E_{\rm T}$ in the Breit
system in order to avoid regions of the dijet phase space which are sensitive
to soft gluon radiation. The data (fig.~\ref{fig10}) are compared with DISENT
NLO calculations, which have been checked with MEPJET. For $O(10)$ GeV$^2 < Q^2
< O(10.000)$ GeV$^2$ and $E_{{\rm T}B} > 5$ GeV, i.e. where NLO and
hadronization corrections are small, the studied jet observables are reproduced
within about 10\%; for $Q^2 < 10$ GeV$^2$, however, where NLO corrections are
becoming large, significant disagreement is observed. (For new data on multijet
production by ZEUS see Chap.~2.3.)

From the dijet fraction $R_{2+1}(Q^2)$ measured by ZEUS, $\alpha_s$ was
determined using the same method as applied to inclusive jets (Chap. 2.1) with
the result $\alpha_s(M_Z) = 0.1166 \pm 0.0019 \rm{(stat)} +0.0024/0.0033
\rm{(exp)} + 0.0057/-0.0044 \rm{(theor)}$. Also the dependence of $\alpha_s$ on
the scale $Q$ is in good agreement with expectations.

\begin{figure}[ht]
\begin{center}
\epsfig{file=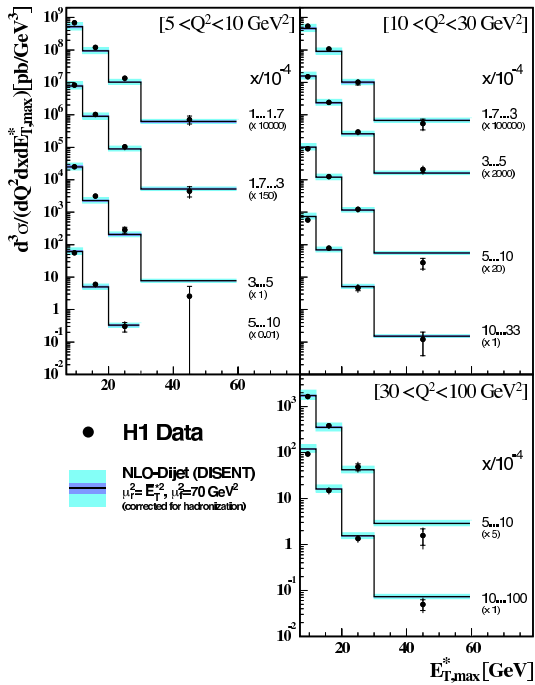,width=7cm}
\caption{\small Inclusive dijet cross section in DIS for
$10^{-4} < x < 10^{-2}$ as a function of the maximum transverse jet energy
$E_{\rm Tmax}^*$ in the hadronic cms compared to NLO QCD predictions (for
CTEQ5M) corrected for hadronization. Outer light error band includes
hadronization (dark band) and renorm. scale uncertainties.}
\label{fig11}
\end{center}
\end{figure}

\begin{figure}[ht]
\begin{center}
\epsfig{file=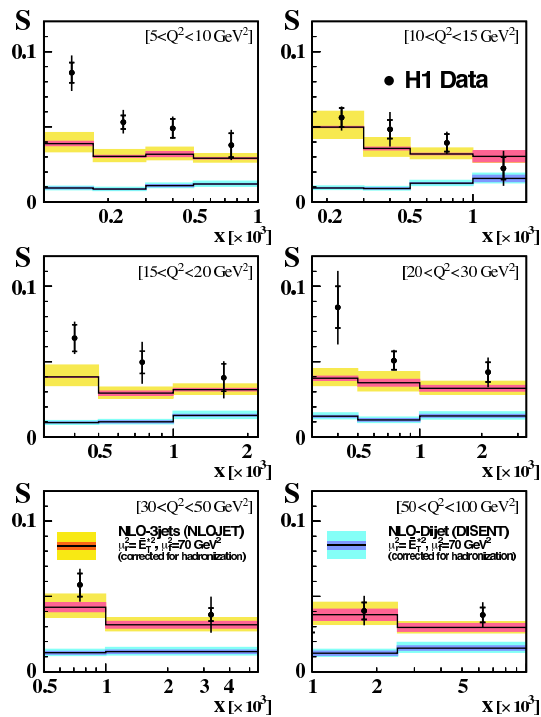,width=7cm}
\caption{\small Ratio $S$ of number of dijet events with small
azimuthal jet separation ($\phi < 2\pi/3$) to the total number of inclusive
dijet events as a function of (Bjorken)$x$ compared to NLO QCD predictions and
with error bands as in fig.~\ref{fig12}.}
\label{fig12}
\end{center}
\end{figure}

Recently, low $x$ dijet production has been studied in more detail by H1
\cite{r9} in the kinematic region 5 GeV$^2 < Q^2 < 100 $ GeV$^2$, $10^{-4} < x
< 10^{-2}$ and $0.1 < y < 0.7$. For jets reconstructed in the hadronic cms with
cuts of $E_{\rm T1} > 7$ GeV, $E_{\rm T2} > 5$ GeV within $-1 < \eta < 2.5$,
the triple differential dijet cross sections in ($x, Q^2, E_{\rm Tmax}$)
(fig.~\ref{fig11}) and ($x, Q^2, \Delta \eta$) do not show significant
deviations from NLO calculations with $\mu_R = (E_{\rm T1} + E_{\rm T2})/2$,
$\mu_F = \sqrt{\langle E_{\rm T}^2 \rangle} = \sqrt{70 {\rm GeV}^2}$, in
contrast to the ratio of the number of dijet events with azimuthal angle
separation of less than 120$^{\circ}$ to all dijet events (fig.~\ref{fig12}).
Better agreement for part of the phase space can be achieved by
combining LO ME with direct photon processes and $k_T$ ordered parton emission.

\subsection{Trijets}

While the cross sections for inclusive jet and dijet production in LO pQCD
depend on $O(\alpha_s)$ only, the trijet cross section already in LO is
sensitive to $O(\alpha_s^2)$.

\begin{figure}[ht]
\begin{center}
\epsfig{file=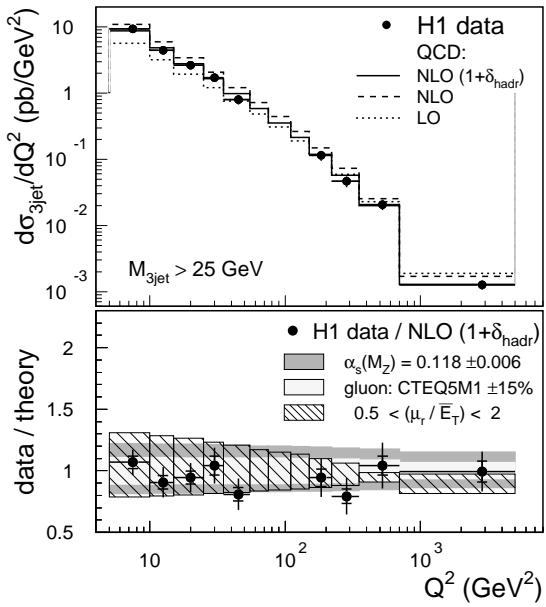,width=7.5cm}
\caption{\small Inclusive trijet production in DIS for 5 GeV$^2  <
Q^2 < 5000$ GeV$^2$; $-1 < \eta_{\rm lab} < 2.5; M_{3j} > 24$ GeV. QCD:
NLOJET.}
\label{fig13}
\end{center}
\end{figure}

Trijet production in DIS has been measured by H1 \cite{r10} for 5 GeV$^2  < Q^2
< 5000$ GeV$^2$, selecting jets with $E_{\rm T} > 5$ GeV in the Breit frame.
The measured cross sections (fig.~\ref{fig13}) were compared with LO and NLO
calculations using NLOJET with $\mu_R$ and $\mu_F$ set to the average
transverse energy $\overline{E_{\rm T}}$ of the 3 jets in the Breit frame; for
the proton PDFs CTEQ5M1 was taken with $\alpha_s(M_Z)=0.118$. The inclusive
trijet cross sections for invariant masses $> 25$ GeV as well as the
trijet/dijet ratio, in which many theoretical and experimental uncertainties
cancel out,  agree well with the NLO predictions in spite of the large NLO
corrections. Also the angular distribution of trijets agrees well with the pQCD
expectation.

\begin{figure}[ht]
\begin{center}
\epsfig{file=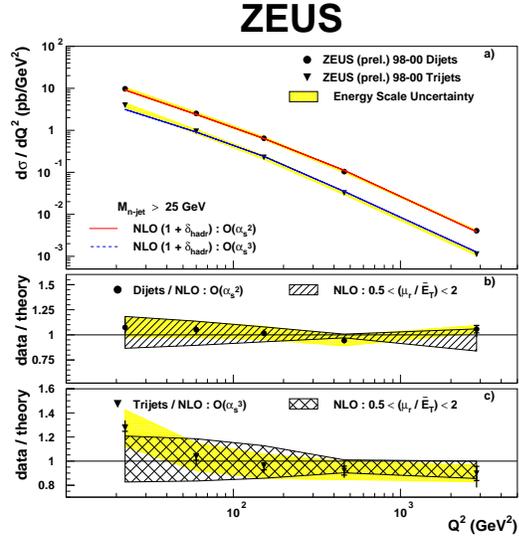,width=7.5cm}
\caption{\small Inclusive dijet and trijet cross section in
DIS. The light shaded band indicates the calorimeter energy scale uncertainty,
the hatched band the renormalization scale uncertainty.}
\label{fig14}
\end{center}
\end{figure}

\begin{figure}[ht]
\begin{center}
\epsfig{file=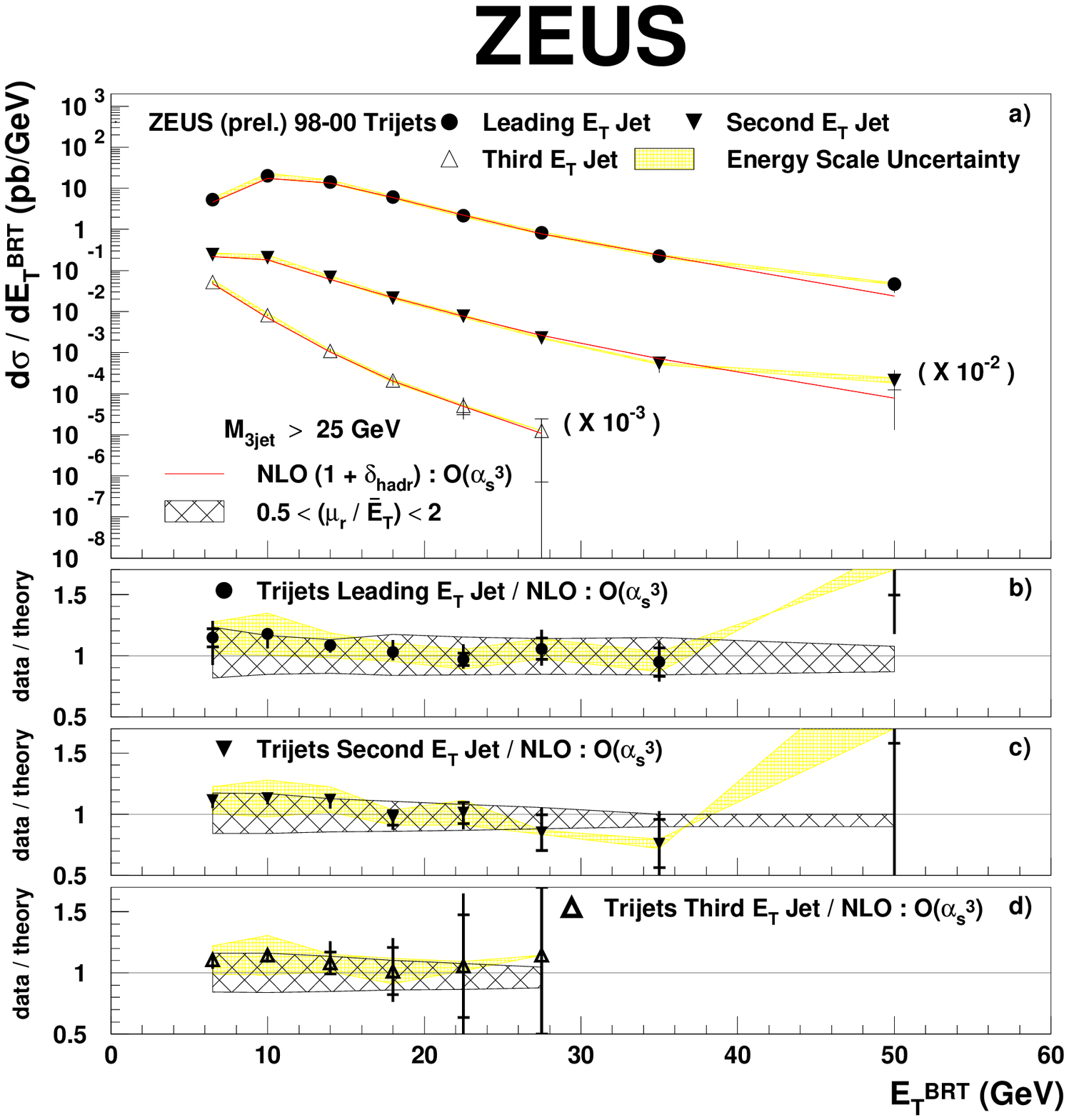,width=7.5cm}
\caption{\small Inclusive trijet cross section in DIS; error
bands as in fig.~\ref{fig14}.}
\label{fig15}
\end{center}
\end{figure}

Recent studies of dijet and trijet production by ZEUS \cite{r11} for 10 GeV$^2
< Q^2 < 5000$ GeV$^2$ in the kinematic range $0.04 < y < 0.6$, $\cos
\gamma_{\rm hadr} < 0.7$, with cuts in the jet phase space of $-1 < \eta_{\rm
lab} < 2.5$, $E_{{\rm T}B} > 5$ GeV for invariant masses of $M_{2j, 3j} > 25$
GeV are shown in figs.~\ref{fig14}~and~\ref{fig15}. Good agreement is found
with the pQCD predictions after correcting for hadronization, i.e. to
$O(\alpha_s^3)$. 

\subsection{Subjet multiplicities}

Reapplication of the jet algorithm with smaller resolution scale $y_{cut}$ to
identified jets allows the resolution of clusters within the jets which are
called subjets. The development of the subjet multiplicity with $y_{cut}$ can
be calculated in pQCD for high $E_{{\rm T}j}$ and not too low $y_{cut}$ 
values, i.e. where fragmentation effects are estimated to be small in
NLO. Since the subjet multiplicity is mainly determined by QCD radiation in the
final state, it only weakly depends on the PDFs of the proton.

\begin{figure}[ht]
\begin{center}
\epsfig{file=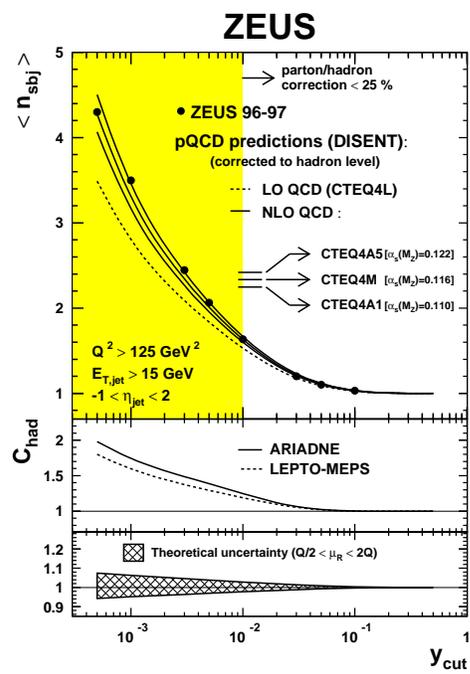,width=7.5cm}
\caption{\small Mean subjet multiplicity as a function of
$y_{cut}$ in inclusive jet production in DIS for $E_{{\rm T}j} > 15$ GeV, $Q^2
> 125$ GeV$^2$ and $-1 < \eta_j < 2$.}
\label{fig16}
\end{center}
\end{figure}

The mean subjet multiplicity has been measured by ZEUS \cite{r12} in NC DIS
processes with $Q^2 > 125$ GeV$^2$ for $E_{\rm T} > 15$ GeV and $-1 < \eta_j <
2$. The ILICA was applied to jets in the lab frame, where calculations of up to
$O(\alpha_s^2)$ can be performed. The mean subjet multiplicity $<n_{sbj}>$ was
determined for $5.10^{-4}< y_{cut} < 0.1$. The data are in good agreement with
NLO calculations with $\mu_R = \mu_F = Q$ using the CTEQ4M PDFs.
(fig.~\ref{fig16}).

Using the method described earlier (see Chap. 2.1), $\alpha_s$ was determined
from these data for 25 GeV$ < E_{\rm T} < 71$ GeV and $y_{cut} =
10^{-2}$ yielding $\alpha_s(M_Z) = 0.1187 \pm 0.0017 \rm{(stat)} +0.0024/-0.0009 \rm{(syst)} 
+0.0093/-0.0076 \rm{(theor)}$. While the experimental uncertainty is comparable
to that resulting from jet measurements, the theoretical uncertainty is larger
and dominated by NNLO terms.

\section{Jets in high energy photoproduction}

In LO QCD direct as well as resolved processes (c.f. Chap.~1) have to be
considered in the calculation of the jet cross section. If $x_\gamma$ is the
fractional momentum of parton $a$ of the photon and $x_p$ the same of parton
$b$ of the proton, the jet cross section is obtained from the convolution of
the photon PDF $f_{\gamma b}(x_\gamma, \mu^2)$ and the proton PDF $f_{pa}(x_p,
\mu^2)$ with the hard partonic cross section $d\sigma_{a,b}(x_{\gamma}, x_p,
\alpha_s, \mu^2)$

\begin{eqnarray} 
\sigma_j & = & \sum_{a,b} \int \int{dx_{\gamma}}dx_p f_{pa}(x_p, \mu^2)
f_{\gamma_b}(x_{\gamma},\mu^2) \nonumber \\ 
& & \cdot d\sigma_{a,b}(x_{p},
x_{\gamma}, \alpha_s, \mu^2) \cdot (1+\delta_{\rm hadr}) 
\end{eqnarray}

\noindent summed over all partons $a,b$ from the photon and proton. For direct
photon interactions $f_{\gamma}$ is a $\delta$-function at $x_{\gamma} = 1$.
Since the transverse energy $E_{\rm T}$ of the jet or jet system is the only
scale parameter available in the photoproduction of jets from light quarks,
$\mu_R$ and $\mu_F$ are set to $\mu_R = \mu_F = \mu = E_{\rm T}$. In resolved
processes photon fragments may interact with proton remnants causing a
background of multiparton interactions ("underlying events"), which has to be
corrected for; it remains below 10\% for $E_{\rm T} > 20$ GeV.

At HERA, the proton PDFs are, with the current statistics, predominantly probed
in the range $0.05 < x_p < 0.5$, where they are well constrained by
measurements of $F_2^p$. For the photon PDFs the probed range is $0.1 <
x_{\gamma} < 1$ in which quarks are well constrained for $x_{\gamma} < 0.5$ by
measurements of $F_2^{\gamma}$ at LEP, while the gluon is only poorly
constrained. Since jet production at HERA is sensitive to the gluon component 
in the photon at LO, it should allow to extend the study of the photon
structure to higher scales.

In single jet production, in comparison to dijet production, an increased
kinematic range is accessible, higher statistical accuracy can be achieved and
infrared dangerous regions of phase space are not existent; on the other hand
single jet photoproduction is less sensitive to details of the hard scattering
process.

\subsection{Inclusive jets}

\begin{figure}[ht]
\begin{center}
\epsfig{file=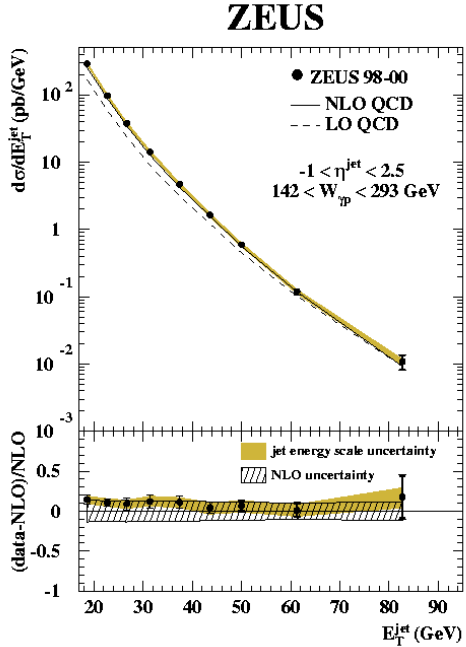,width=7cm}
\caption{\small Inclusive jet production in photoproduction ($Q^2 < 1$ GeV$^2$)
integrated over $-1 < \eta_j < 2.5$. LO and NLO: Klasen, Kleinwort, Kramer with
$\mu_R = \mu_F = E_{\rm T}$, proton-PDFs: MRST99, photon-PDFs: GRV.}
\label{fig17}
\end{center}
\end{figure}

H1 \cite{r13} and ZEUS \cite{r14} have performed similar measurements and
analyses of inclusive jet cross sections for photons of $Q^2 < 1$ GeV$^2$
within $-1 < \eta_{j} < 2.5$. The H1 data for $E_{\rm T} > 21$ GeV cover a
range of the hadronic cms-energy $W_{\gamma p}$  of 95 GeV $< W_{\gamma p} <
285$ GeV. The measurements were extended down to $E_{\rm T} > 5$ GeV with a
dedicated trigger for photons of $Q^2 < 0.01$ GeV$^2$ in the restricted
kinematic range of 164 GeV $< W_{\gamma p} < 242$ GeV. The ZEUS data were taken
for 142 GeV $< W_{\gamma p} < 293$ GeV for $E_{\rm T} > 17$ GeV (see
fig.~\ref{fig17}). H1 compares the data with the NLO QCD calculations of
Frixione and Ridolfi, using CTEQ5M1 for the proton PDFs and GRV for the photon
PDFs and setting $\mu_R = \mu_F$ to the average transverse energy of the two
outgoing partons; ZEUS compares with Klasen, Kleinwort, Kramer, using MRST99
for the proton and GRV for the photon with $\mu_R = \mu_F = E_{{\rm T}j}$. In
both analyses good agreement with NLO QCD over six orders of magnitude of the
cross section is found; the H1 data for $Q^2 < 0.01$ GeV and 5 GeV $< E_{\rm T}
< 12$ GeV may indicate a cross section rising above prediction with increasing
$\eta$.

From the ZEUS data, $\alpha_s(M_Z)$ was determined (see Chap.~2.1) by
calculating in NLO QCD for each $E_{\rm T}$-bin $i$ the cross section
$(d\sigma/dE_{\rm T})_i$ using the three sets of MRST99 with their
corresponding $\alpha_s(M_Z)$-values. By combining the $\alpha_s$ values for
the different $E_{\rm T}$ bins a value of $\alpha_s = 0.1224 \pm 0.0002
\rm{(stat)} +0.0022/-0.0019 \rm{(exp)} + 0.0054/-0.0019 \rm{(theor)}$ was
obtained. The major error sources are the jet energy scale ($\pm 1.5$ \% in
$\alpha_s(M_Z)$) and the renormalization scale ($+4.2/-3.3$\% in
$\alpha_s(M_Z)$); the uncertainties from the PDFs and hadronization are
estimated to be less than 1\%. 

Similarly, the dependence of $\alpha_s$ on the energy scale was investigating
by determining $\alpha_s$ from $d\sigma/dE_{\rm T}$ at different $E_{\rm T}$
values. The resulting $\alpha_s (E_{\rm T})$ values agree well with the
predicted running of $\alpha_s$. 

\begin{figure}[ht]
\begin{center}
\epsfig{file=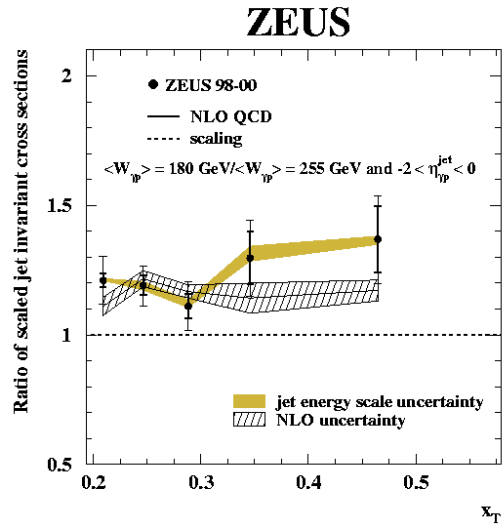,width=7cm}
\caption{\small Scaling violation in photoproduction ($Q^2 < 1$ GeV$^2$):
Ratio of measured scaled invariant jet cross sections (see text) for the two
$\langle W_{\gamma p}\rangle$ shown and averaged over $-2 < \eta_j^{\gamma p} <
0$. NLO QCD as in fig.~\ref{fig17}.}
\label{fig18}
\end{center}
\end{figure}

Using the above data, ZEUS has measured the dependence of the scaled invariant
jet cross sections $E_{{\rm T}j}^4 E_{j}d^3 \sigma / dp_{xj} dp_{yj} dp_{zj}$
on the dimensionless variable $x_{\rm T} = 2 E_{{\rm T}j}/W_{\gamma p}$. The
cross section measured at \mbox{$\langle W_{\gamma p} \rangle $} of 180 and 255
GeV and averaged over $-2 < \eta_j^{\gamma p} < 0$, agrees well with NLO QCD.
In the quark parton model, the scaled cross section should be independent of
$W_{\gamma p}$, whereas from QCD one expects a violation of scaling due to the
evolution of the structure functions and the running of $\alpha_s$. As shown in
fig.~\ref{fig18}, the ratio of the scaled cross section at the two different
energies $W_{\gamma p}$ violates scaling as predicted by NLO QCD.

\subsection{Inclusive dijets}

Dijet production has been measured by ZEUS \cite{r15} and H1 \cite{r16} under
similar kinematical conditions as in single jet production. In both experiments
the jets were reconstructed with the ILICA, requiring asymmetric cuts on
$E_{\rm T}$ in the lab frame. Both experiments cover a range of $-1 <
\eta_{j1,2} < 2.5$ and $0.1 < y < 0.9$. The quantity $x^{obs}_{\gamma}$ is
introduced

\begin{equation}  x^{obs}_{\gamma} = (E_{{\rm T}j1}e^{-\eta_{1}}+E_{{\rm
T}j2}e^{-\eta_{2}})/2yE_e \end{equation}

\noindent where $y = W^2/s$, which in LO corresponds to the fraction of the
photon momentum that contributes to the production of the two highest-$E_{\rm
T}$ jets; cuts on $x^{obs}_{\gamma}$ allow the data sample to be enriched with
direct processes (high $x_\gamma^{obs}$) or resolved processes (low
$x_\gamma^{obs}$). The measurements are compared to NLO QCD calculations with
$\mu_R = \mu_F$ equal set to the average $E_{\rm T}$ of the two outgoing
partons. The PDF-parametrizations used are GRV-HO and AFG-HO for the photon and
CTEQ5M1 for the proton; $\alpha_s(M_Z)$ is set to 0.118.


\begin{figure}[ht]
\begin{center}
\epsfig{file=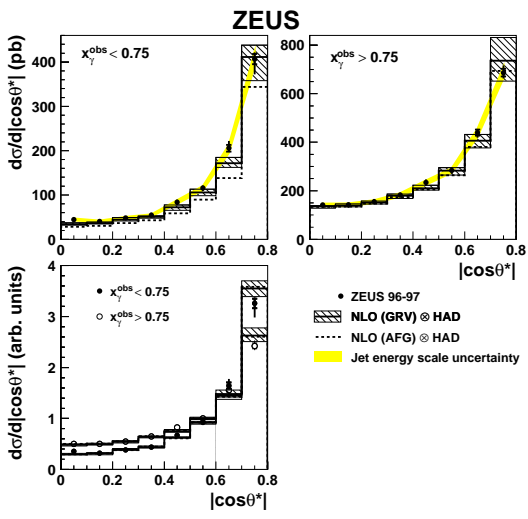,width=7cm}
\caption{\small Inclusive dijet production in photoproduction ($Q^2 < 1$
GeV$^2$) as a function of the dijet angle $\Theta^*$ in the parton-parton-cms
for different cuts on the fractional momentum $x_\gamma^{\rm obs}$ of the
photon participating in the production of the two jets with the highest
transverse energy. Cuts: $E_{{\rm T}j1,2} > 14, 11$ GeV; $-1 < \eta_j < 2.4;
134 < W_{\gamma p} < 277$ GeV. NLO: Frixione, Ridolfi with CTEQ5M for
proton-PDFs and GRV resp. AFG for photon-PDFs.}
\label{fig19}
\end{center}
\end{figure}

\begin{figure}[p]
\begin{center}
\epsfig{file=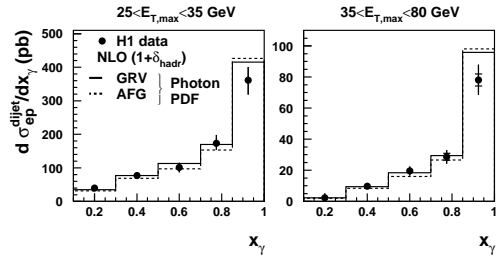,width=7cm}
\caption{\small Inclusive dijet production in photoproduction ($Q^2 < 1$
GeV$^2$) as function of the fractional momentum $x_{\gamma}$ of the photon
participating in the production of the two jets with highest transverse energy
$E_{\rm T}$ (identical to $x_\gamma^{obs}$ in fig.~20). Cuts: $E_{{\rm T}j1,2}
> 25, 15$ GeV; $-0.5 < \eta_j < 2.5$; $95 < W_{\gamma p} < 285$ GeV. QCD, PDFs
as in fig.~\ref{fig19}.}
\label{fig20}
\end{center}
\end{figure}

The measured angular distributions (fig.~\ref{fig19}) confirm the steeper rise
expected for resolved photon processes $(x_\gamma^{obs} < 0.75)$ in comparison
to direct photon processes $(x_\gamma^{obs} > 0.75)$, indicating that the
dynamics of the hard scattering process is reasonably well described. For the
H1 data, the cross section as a function of $x_\gamma$ (here identical to
$x_\gamma^{obs}$), shown in fig.~\ref{fig20} for two regions of $E_{{\rm
T},max}$, which represent two different factorization scales for the PDFs of
the proton and photon, agrees with NLO predictions and varies only slightly
with the different parametrizations of the photon PDFs while ZEUS observes
differences; it has to be noted, however, that the different $E_{\rm T}$-cuts
have been used (H1: $E_{\rm T1} > 25$ GeV, $E_{\rm T2} > 15$ GeV; ZEUS: $E_{\rm
T1} > 14$ GeV, $E_{\rm T2} > 11$ GeV).

\section{Summary of results on $\alpha_s$}

\begin{figure}[p]
\begin{center}
\epsfig{file=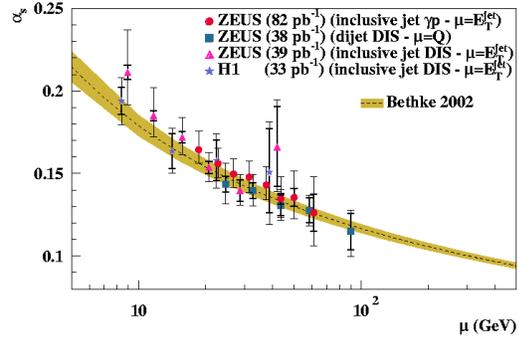,width=7cm}
\caption{\small Scale dependence of $\alpha_s$ from recent
HERA experiments.}
\label{fig21}
\end{center}
\end{figure}

\begin{figure}[p]
\begin{center}
\epsfig{file=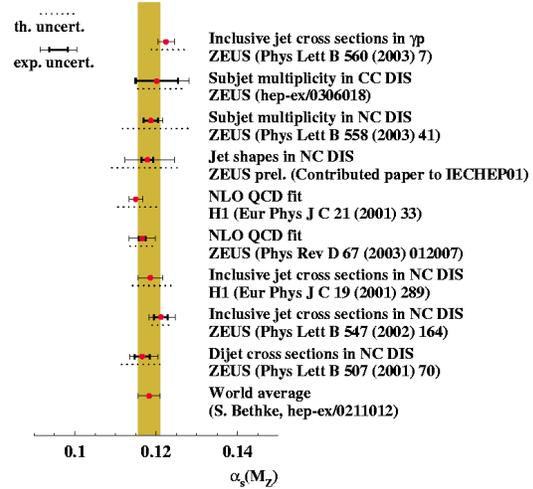,width=7cm}
\caption{\small Determination of $\alpha_s(M_Z)$ and its scale
dependence in recent HERA experiments.}
\label{fig22}
\end{center}
\end{figure}

The results obtained from the recent measurements of jet and subjet cross
sections in DIS and photoproduction on $\alpha_s(M_Z)$ as well as on the scaling behaviour
of $\alpha_s$ have achieved an accuracy compatible with results from other
colliders. They are also in good agreement with the world average
(figs.~\ref{fig21} and \ref{fig22}).

\section{Summary}

For large scales, $Q^2$ and/or jet-$E_{\rm T}$, the recent results from H1 and
ZEUS on jet production in deep inelastic scattering and photoproduction are in
good agreement with NLO QCD. In general, the hadronization corrections applied
to the NLO results are small and improve the description of the data only
slightly. For decreasing scales and in specific areas of phase space the
corrections become large and are important in improving the description of the
data.

In these studies, the longitudinally invariant $k_T$ cluster algorithm in its
inclusive mode has proved reliable and, therefore, become the preferred method
of jet identification in this field. The good agreement of $\alpha_s$
determinations performed with jet (and subjet) production at high $Q^2$ and 
$E_{\rm T}$ with the world average supports this conclusion.

The situation is worse in the small $x$ / forward region. For $x < 10^{-4}$ and
small $Q^2$ a breakdown of DGLAP is to be expected and indeed increasing
discrepancies with DGLAP based predictions are observed even in NLO. In some
cases in spite of small experimental errors the sizable theoretical
uncertainties, however, do not yet allow safe conclusions on a more
satisfactory and comprehensive description of the data by modified pQCD models
like BFKL or CCFM.

Not only an improvement of experimental statistical and systematic errors, to
come from HERA II, but also more accurate model calculations are needed to get
closer to a satisfactory understanding of the interplay of soft and hard QCD
processes in DIS and photoproduction. 

\vskip1cm

{\underline {Acknowledgement}}:

\vskip5pt

My thanks are due to P.N.~Bogolyubov and L.~Jenkovsky for the invitation to an
interesting and enjoyable conference. I am indebted to G.~Grindhammer for
valuable discussions and to him as well as to T.~Greenshaw for a critical
reading of the paper. I also thank Kristiane Preuss and Marlene Schaber for
their technical help with the paper.

\end{document}